\documentclass[final]{elsarticle}

\usepackage{lineno,hyperref,amssymb}

\journal{Acta Materialia}

\begin{document}

\begin{frontmatter}

\title{Thermodynamic investigations on the growth of CuAlO$_2$ delafossite crystals}

\author{Nora Wolff\corref{mycorrespondingauthor}}
\cortext[mycorrespondingauthor]{Corresponding author}
\ead{nora.wolff@ikz-berlin.de}

\author{Detlef Klimm\corref{}}
\author{and Dietmar Siche\corref{}}

\address{Leibniz-Institut f\"ur Kristallzu\"uchtung, Max-Born-Str. 2, 12489 Berlin, Germany}

\begin{abstract}
Simultaneous differential thermal analysis (DTA) and thermogravimetric (TG) measurements with copper oxide/aluminum oxide mixtures were performed in atmospheres with varying oxygen partial pressures and with crucibles made of different materials. Only sapphire and platinum crucibles proved to be stable under conditions that are useful for the growth of CuAlO$_2$ delafossite single crystals. Then the ternary phase diagram Al$_2$O$_3$--CuO--Cu and its isopleth section Cu$_2$O--Al$_2$O$_3$ were redetermined. Millimeter sized crystals could be obtained from copper oxide melts with 1--2\,mol-\% addition of aluminum oxide that are stable in platinum crucibles held in oxidizing atmosphere containing 15--21\% oxygen.
\end{abstract}


\end{frontmatter}



\section{Introduction}

The technological interest in developing transparent conducting oxide (TCO) or transparent semiconducting oxide (TSO) materials has seen a dramatic increase in recent years. Established TCO's, such as CdO, $\beta$-Ga$_2$O$_3$, In$_2$O$_3$, SnO$_2$ and ZnO \cite{Galazka15,Klimm14b} are n-type semiconductors in which electrons act as main charge carriers. However, in semiconductor electronics, p-n-junctions are of major importance where hole conduction is required in the p-type parts \cite{Schulz12}. P-type TCO's could find their applications in functional p-n heterojunctions for transparent solar cells, ultraviolet-emitting diodes (LED) \cite{Kawazoe97,Banerjee07}, touch screens, thermoelectric converters and other transparent optoelectronic devices \cite{Liu14}. Unfortunately the valence bands of most oxides are very flat. This impedes high hole mobility which would be important for commercial devices. Some copper(I) compounds, including the delafossite type CuAlO$_2$ were identified belonging to the small group of prospective p-type TCO's \cite{Hautier13}.

In the ternary delafossite oxide structure of CuAlO$_2$ $\lbrace$O---Cu---O$\rbrace$ dumbbell layers are stacked with {AlO$_6$} octahedra and hexagonal Cu layers perpendicular to the c axis \cite{Lee01}. CuAlO$_2$ delafossite has trigonal symmetry (space group $R\bar{3}m$), with $a=b=2.857$\,\AA, $c=16.939$\,\AA, an indirect bandgap of 2.22\,eV and a direct gap of 3.4\,eV \cite{Yoon13}.

So far, CuAlO$_2$ has mainly been grown as polycrystalline thin layers or on a ceramic route \cite{Kawazoe97,Yanagi00,Byrne14}. Only scarce reports on the growth of small single crystal can be found in the literature \cite{Lee01,Yoon13,Ishiguro81,Brahimi14} and all of them rely on crystallization from melt solutions based on excess copper oxides (Cu$_2$O, CuO) and Al$_2$O$_3$. This results in millimeter-sized (0001) platelets with several 100\,$\mu$m thickness. It is surprising that on the one hand recent authors used often both Cu$_2$O and CuO as components of the flux, but on the other hand not much attention is given to the growth atmosphere. Via the reaction 
\begin{equation}
\mathrm{Cu}_2\mathrm{O} + \frac{1}{2}\mathrm{O}_2 \rightleftarrows 2\,\mathrm{CuO} + \it\Delta H   \label{eq:Cu-O}
\end{equation}
both copper oxides are in equilibrium. The enthalpy of the exothermal oxidation (\ref{eq:Cu-O}) as calculated with the FactSage integrated thermodynamic databank system \cite{FactSage7_1} changes almost linearly from $-142$\,kJ/mol at $0^{\,\circ}$C to $-129$\,kJ/mol at $1200^{\,\circ}$C which means that high temperature $T$ and low oxygen fugacity (expressed as partial pressure $p_{\mathrm{O}_2}$) shift (\ref{eq:Cu-O}) to the educt side with copper(I) oxide.

Over a wide range of experimental conditions $(T, p_{\mathrm{O}_2}) $, covering all conditions that are used in this study, aluminum occurs exclusively as Al$^{3+}$, and hence Al$_2$O$_3$ can be considered as a component of the system. 2\,CuAlO$_2=$\,Cu$_2$O$\cdot$Al$_2$O$_3$ is the 1:1 intermediate compound in the pseudobinary system Cu$_2$O--Al$_2$O$_3$. However, via (\ref{eq:Cu-O}) CuO will be present in the system, and can form with Al$_2$O$_3$ the spinel phase CuAl$_2$O$_4$ \cite{ONeill05}. It will be shown later that under certain experimental conditions also metallic copper can be formed, and consequently the concentration triangle Al$_2$O$_3$--CuO--Cu in Fig.~\ref{fig:triangle}a) will be used in this paper to discuss growth conditions for CuAlO$_2$ delafossite crystals.

In a report on the separate systems CuO--Al$_2$O$_3$ and Cu$_2$O--Al$_2$O$_3$ it was found that the spinel CuAl$_2$O$_4$ is stable in air only up to $1000^{\,\circ}$C and is then converted to CuAlO$_2$, which was stable to about $1260^{\,\circ}$C \cite{Misra63}. Gadalla and White \cite{Gadalla64} reported the ternary Al$_2$O$_3$--CuO--Cu$_2$O for oxygen fugacities $0.21\leq p_{\mathrm{O}_2}/\mathrm{bar}\leq1.0$. According to their data, single crystals of CuAlO$_2$ should crystallize from melts containing 10-20\,mol\% Al$_2$O$_3$ in Cu$_2$O, which is in slight contrast with the results that will be presented in this study. Nevertheless, their study proposes for $p_{\mathrm{O}_2}=0.21$\,bar peritectic melting of CuAlO$_2$ at $1238^{\,\circ}$C to $\alpha$-Al$_2$O$_3$ and a melt consisting mainly of CuO and Cu$_2$O, which is in fairly good agreement with the results that will be presented here. For large $p_{\mathrm{O}_2}\approx0.4$\,bar, the stability range for the CuAl$_2$O$_4$ spinel was found elsewhere \cite{Jacob75} to be wider, compared to Gadalla and White \cite{Gadalla64}, but this difference is not relevant for the growth of the copper(I) compound CuAlO$_2$.

It is the aim of this work to define experimental conditions (melt composition, atmosphere, temperature regime, crucible material) where CuAlO$_2$ single crystals can be growth from self-flux. Self-fluxes are expected to result in crystals with less impurities, compared to growth from fluxes containing foreign components, such as hydrothermal solutions, or by sol-gel processes \cite{Sato08,Shahriari01,Gotzendorfer10}.


\section{Experimental}

Simultaneous differential thermal analysis (DTA) and thermogravimetric (TG) measurements were carried out with a Netzsch STA 409CD. Gas flow was set by a mass flow controller do deliver atmospheres ranging from 100\% O$_2$ to 2\% O$_2$ in Ar, and air. Samples were prepared by mixing powders of Al$_2$O$_3$ (Alfa Aesar, 99.997\% purity) and Cu$_2$O (Fox Chemicals, 99.99\% purity) to concentrations between pure Cu$_2$O and 15\,mol-\% Al$_2$O$_3$ in Cu$_2$O. For all measurements a temperature program was selected, which resembles the conditions during a real crystal growth run. With this program, it was possible to test potential crucible materials for their steadiness and to obtain some first information about redox processes of copper oxides. The powder mixtures were heated in crucibles with a rate of 5\,K/min to 1200 or $1300^{\,\circ}$C (3\,h dwell time) then slowly cooled down with 3\,K/min to $900^{\,\circ}$C and 1.5\,K/min to $100^{\,\circ}$C. The dwell time of 3\,h was chosen to test the stability of the respective crucible material. In order to ensure that the CuAlO$_2$ phase was formed, the temperature program was run twice.

The phase compositions of all samples were characterized by X-ray powder diffraction with a GE Inspection Technologies XRD 3003 TT (Bragg Brentano geometry and Cu K$\alpha$ radiation).

The CuAlO$_2$ composition of the crystals was proven by X-ray fluorescence spectroscopy.


\section{Results and discussion}

\begin{figure}[htb]
\includegraphics[width=0.60\textwidth]{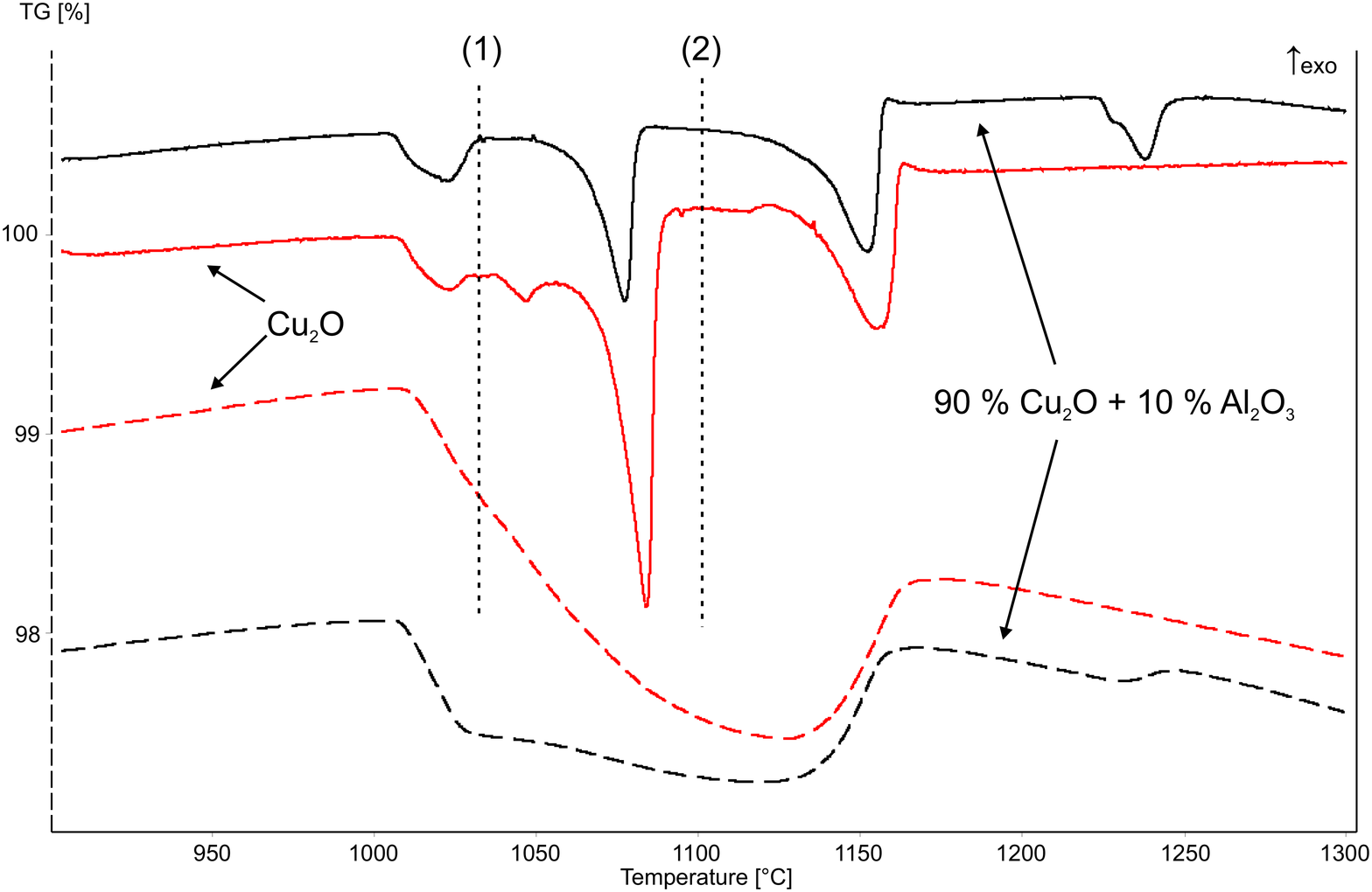}
\caption{DTA (solid) and TG curves (dashed) between $900^{\,\circ}$C and $1300^{\,\circ}$C for pure Cu$_2$O and a 90/10 mixture with Al$_2$O$_3$. The DTA peaks between $1000^{\,\circ}$C and $1150^{\,\circ}$C are overlapped by redox processes (\ref{eq:Cu-O}) (mass change in TG curve).}
\label{fig:2samples}
\end{figure}

\begin{figure}[htb]
\includegraphics[width=0.60\textwidth]{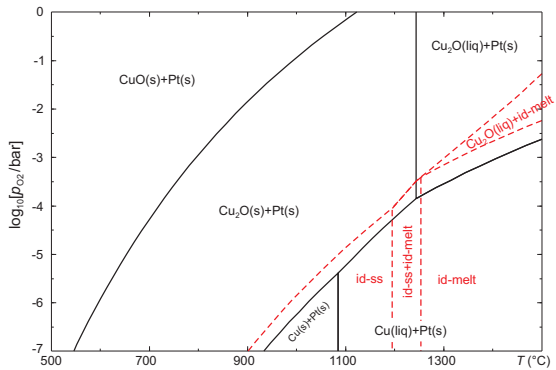}
\caption{Predominance diagram of the system Cu--Pt-O$_2$ in coordinates $T,\log[p_{\mathrm{O}_2}]$. Solid black lines: assumption of pure phases CuO(s), Cu$_2$O(s,liq), Cu(s,liq), Pt(s). Dashed red lines: Pt and Cu are assumed to form ideal solutions in the solid and liquid phases (id-ss and id-melt, respectively).}
\label{fig:predominance}
\end{figure}

First measurements gave information on the oxidation and reduction behavior of copper(I)oxide. If Cu$_2$O (pure or mixed with Al$_2$O$_3$) was heated in an atmosphere containing O$_2$, oxidation to CuO occured around $350^{\,\circ}$C, which is in agreement with Gadalla and White \cite{Gadalla64}.
DTA-peaks between $1000^{\,\circ}$C and $1050^{\,\circ}$C result from endothermal melt processes and also from endo- or exothermal redox processes between Cu$^+$ and Cu$^{2+}$ (\ref{eq:Cu-O}). An oxidation process is connected with a mass increase in the associated TG curve, together with an exothermal DTA effect. A reduction is connected with a mass decrease because of oxygen release, and is endothermal. This is demonstrated by the red curves in Fig.~\ref{fig:2samples} which were obtained with pure Cu$_2$O. The first three endothermal peaks are connected with a mass loss that is monotonous but changes in rate: E.g. at point (1) where the DTA curve is less endothermal, also the TG curve is flatter. Obviously the reaction rate is smaller at this point, and consequently mass loss is reduced, in agreement with a smaller endothermal effect. The reduction process CuO $\rightarrow$ Cu$_2$O is completed after the large sharp peak, at point (2) where the sample is still solid. A small exothermal effect after this point and a mass gain of almost 1\% starting there indicates that Cu$_2$O begins re-oxidizing partially back to CuO. This is surprising because according to Fig.~\ref{fig:predominance} under constant $p_{\mathrm{O}_2}$ a higher $T$ should result in lower rather than in higher valence. Obviously, the energy gain resulting from liquid mixture formation, in analogy to (\ref{eq:G_id}), is the driving force for the re-oxidation. With increasing CuO content the sample can melt, giving rise the the following broad peak. Visual inspection of samples that were heated up to $1200^{\,\circ}$C proved this.

An identical measurement with a Cu$_2$O/Al$_2$O$_3$ mixture (black curves in Fig.~\ref{fig:2samples}) is similar, but shows an additional endothermal peak between $1230^{\,\circ}$C and $1250^{\,\circ}$C. It will be shown later that this peak can be assigned to the peritectic melting of CuAlO$_2$.

In further experiments different crucible materials in combination with varying oxygen fugacities in the atmosphere were tested. Ceramic crucibles, for example Al$_2$O$_3$, Y$_2$O$_3$ \cite{Yamada93} or MgO, were infiltrated by copper oxide flux and could be used only in the low-$T$ range where the whole sample remains solid. This is in agreement with a recent report \cite{Diemer99} where wetting of alumina surfaces by CuO$_x$ melts could be reduced only for very low $p_{\mathrm{O}_2}$. In contrast to ceramic Al$_2$O$_3$, single crystalline sapphire crucibles proved satisfactorily stable against melts containing CuO$_x$. After several days, only minor attack to the crucible surface was visible. But if the melt sticks tightly to them, thermo-mechanical shock can lead to cracking upon cooling. Nevertheless, they could be used if handled with care. Unfortunately, gold crucibles that were used up to $975^{\,\circ}$C for the growth of YBa$_2$Cu$_3$O$_{7-x}$ \cite{Tao91} are not a valid alternative because growth temperatures for the current process are too high. Finally, platinum proved to be the most stable crucible material and was used for all further experiments.

\begin{figure}[htb]
\includegraphics[width=0.60\textwidth]{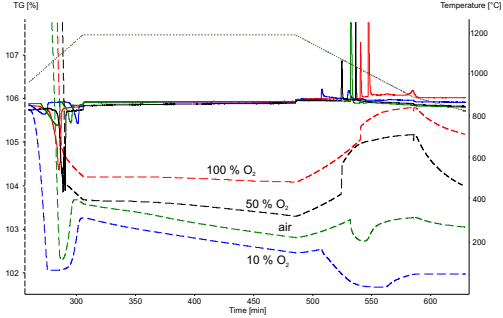}
\caption{TG (dashed) DTA (solid) curves of measurements with different oxygen partial pressure in a time range between heating (from $900^{\,\circ}$C on) and cooling (to $800^{\,\circ}$C). During the dwell time at $1200^{\,\circ}$C the mass loss increases with lower oxygen content (formation of Cu). With higher oxygen content in the atmosphere (here 50\% and 100\%) the spinel type CuAl$_2$O$_4$ crystallizes instead of CuAlO$_2$}
\label{fig:Pt-atmos}
\end{figure}

The behavior of Pt crucibles in relation to the copperoxide-rich melt depends mainly on the oxygen fugacity of the growth atmosphere, as shown in Fig.~\ref{fig:Pt-atmos}. These measurements were performed with identical temperature program but in atmospheres with different $p_{\mathrm{O}_2}$. During the dwell time (3\,h at $1200^{\,\circ}$C) the sample mass decreased with a rate that became larger for smaller $p_{\mathrm{O}_2}$ in the atmosphere, and it will be shown that reduction of copper oxides to metallic copper is the origin of this effect. For $p_{\mathrm{O}_2}<0.15$\,bar the crucibles were destroyed after the experiment by alloy formation of Pt with Cu.

Problems occurring with CuO$_x$ melts in Pt crucibles were already reported by several authors, e.g. \cite{Schultze91,Mudenda14} and are related to the formation of solid solutions between Pt and Cu. Both metals crystallize in $fcc$ structure and the binary phase diagram Cu--Pt is in the high $T$ range $>1000^{\,\circ}$C almost ideal. Only below $\approx880^{\,\circ}$C two intermediate compounds with approximate composition CuPt and Cu$_3$Pt are stable \cite{Abe06}. For the process temperatures of $1200-1300^{\,\circ}$C ($1473-1573$\,K) that were used in this study, one obtains for a solution of two components (such as Cu and Pt) a maximum ideal Gibbs free energy of mixing at a molar fraction $x=0.5$, which amounts to
\begin{equation}
\it\Delta G_\mathrm{id,mix} = RT \,2\,x\ln x = -8.5\ldots9.1\,\mathrm{kJ/mol} \label{eq:G_id}
\end{equation}
($R$ -- gas constant). The energy gain that is given by (\ref{eq:G_id}) results in an unfavored stabilization of solid or liquid intermetallic (Cu/Pt) solutions with respect to CuO$_x$+Pt. This is demonstrated in Fig.~\ref{fig:predominance} where the solid lines describe the stability phase fields of the copper oxides in a Pt crucible. (The diagram was calculated with FactSage \cite{FactSage7_1}. Pt melting at $1770^{\,\circ}$C is out of scale.) The upper border of this figure corresponds to 1\,bar O$_2$, and the figure demonstrates that for low $T$ and high $p_{\mathrm{O}_2}$ CuO(s) is stable in Pt crucibles. Towards the bottom right corner Cu$_2$O and subsequently Cu metal are formed. The vertical straight lines are the melting points of these compounds.

The assumption that Cu(liq) could reside in a Pt(s) crucible (bottom right phase field) is unrealistic; instead an alloy will be formed. Ideal miscibility in both solid and liquid phases results in the phase boundaries that are drawn as dashed red lines. The two vertical dashed lines limiting the ``id-ss+id-melt'' phase field correspond to the solidus and liquidus of the Cu--Pt binary system and are in Fig.~\ref{fig:predominance} ca. 200\,K lower than in the literature \cite{Abe06}, but the width of the two-phase region is realistic. The general trend that phase fields containing pure (not alloyed) Pt(s) will shrink, is not influenced by this difference. One can assume that the identical crystal structure of Cu and Pt, connected with infinite mutual solubility, is the main reason for the instability of CuO$_x$ melt in Pt crucibles. After a sufficiently large number of heating/cooling cycles (ca. 6--8) CuO$_x$ loses its oxygen completely under the formation of (Pt,Cu) alloy, which results finally in crucible destruction. It is especially important that local overheating near the crucible wall is avoided, because the upper boundary of the ``Cu$_2$O+id-melt'' phase field approaches quickly very high $p_{\mathrm{O}_2}$ were the crucible metal cannot be stabilized even under highly oxidizing conditions.

In Fig.~\ref{fig:Pt-atmos} the TG signal shows also mass changes during crystallization near $1000^{\,\circ}$C. For the measurements that were performed in the 10\% O$_2$ atmosphere or in air, a mass loss occurs. This can be attributed to the crystallization of CuAlO$_2$, which is a Cu$^+$ compound --- in contrast to the melt that contains Cu$^+$ and Cu$^{2+}$. For the 50\% and 100\% curves, however, the sample mass increases during crystallization, which results from the formation of CuAl$_2$O$_4$. X-ray powder diffractometry of both types of samples proved the existence of the corresponding copper aluminum oxides. The crossover between spinel and delafossite crystallization is at 25\% O$_2$. 15--21\% O$_2$ are the optimum compromise between crystallization of spinel at too high $p_{\mathrm{O}_2}$, and the formation of too much metallic copper at lower $p_{\mathrm{O}_2}$.

\begin{figure}[htb]
\includegraphics[width=0.60\textwidth]{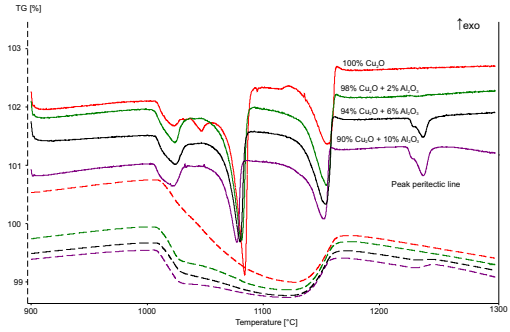}
\caption{From top to bottom: TG (dashed)/DTA (solid) curves of pure Cu$_2$O and of mixtures with increased additions of Al$_2$O$_3$ (measurements in air).}
\label{fig:Al2O3-conc}
\end{figure}

CuAlO$_2$ is reported to melt under peritectic decomposition to $\alpha$-Al$_2$O$_3$ and a melt with copper oxide excess \cite{Misra63,Gadalla64,Jacob75}. For crystal growth experiments it is hence useful, to construct a pseudobinary system Cu$_2$O--Al$_2$O$_3$ and to find the range between the eutectic and peritectic points where delafossite should crystallize first from the melt. This was done in Fig.~\ref{fig:Al2O3-conc} by adding different amounts of Al$_2$O$_3$ to Cu$_2$O. 


For pure Cu$_2$O and the sample with 2\% Al$_2$O$_3$ added the melting process is finished at $1170^{\,\circ}$C. Not so for the samples with 6\% and 10\% Al$_2$O$_3$. Another endothermal peak appears around $T_\mathrm{per}\approx1230-1250^{\,\circ}$C, and this can be assigned to the peritectic melting of CuAlO$_2$. It is interesting to note that also this melting event is connected with a small mass gain, which can be explained as follows: depending on $T$ and $p_{\mathrm{O}_2}$, a certain Cu$_2$O/CuO ratio is in equilibrium. Below $T_\mathrm{per}$, CuAlO$_2$, Cu$_2$O, and CuO are in present. If CuAlO$_2$, which contains only Cu$^+$ melts, more Cu$_2$O is added, and to maintain equilibrium, a certain amount of it must be oxidized to CuO -- which is the origin of the mass gain. 

\begin{figure}[htb]
\includegraphics[width=0.80\textwidth]{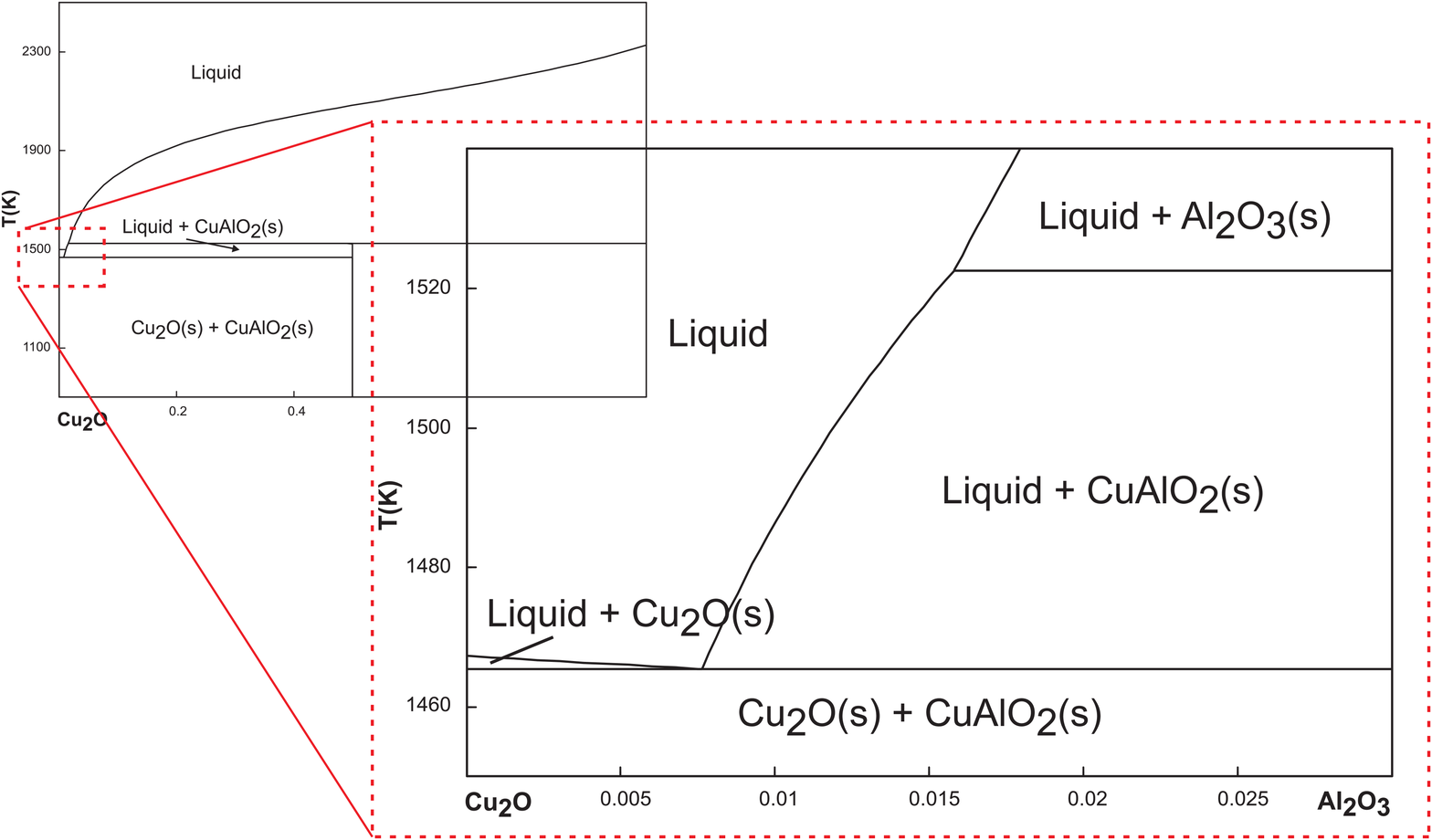}
\caption{Cu$_2$O--Al$_2$O$_3$ pseudobinary section for $p_{\mathrm{O}_2}=0.21$\,bar, with an enlarged image of the CuAlO$_2$ liquidus.}
\label{fig:PD-section}
\end{figure}

\begin{figure}[htb]
\includegraphics[width=0.80\textwidth]{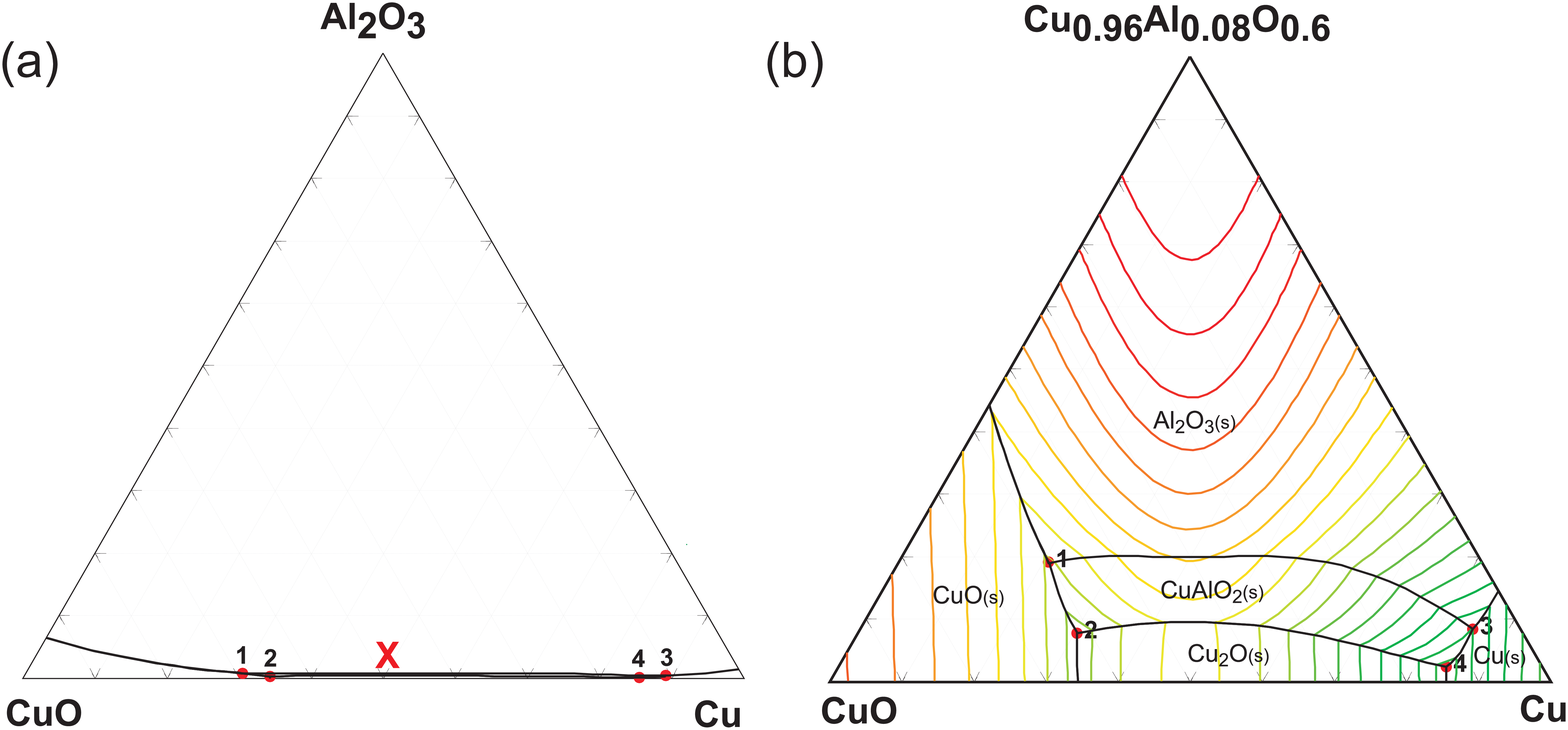}
\caption{Ternary phase diagram Al$_2$O$_3$--Cu--CuO (a) and its lower part Cu$_{0.96}$Al$_{0.08}$O$_{0.6}$--Cu--CuO (b). Isotherms in triangle (b) are drawn every 50\,K; the coldest invariant point 4 is at 1170\,K. The Cu$_{0.96}$Al$_{0.08}$O$_{0.6}$ point is marked as red X in the Al$_2$O$_3$--Cu--CuO system. The labels for all intermediate phases in a) are written in brackets.}
\label{fig:triangle}
\end{figure}

Data from the FactSage \cite{FactSage7_1} database and from the literature \cite{Gadalla64,Schramm05} for CuAlO$_2$, CuAl$_2$O$_4$, CuO(liq), and of the CuO$_x$--Al$_2$O$_3$ melt were refined to describe the experimental findings that are reported here, and the pseudobinary section of the phase diagram was obtained (Fig.~\ref{fig:PD-section}). The eutectic point is extremely near to CuO$_x$, close to 1\% Al$_2$O$_3$. From there the CuAlO$_2$ liquidus spans to ca. 2\% Al$_2$O$_3$, which represents the growth window for this delafossite phase. It should be noted that Fig.~\ref{fig:PD-section} does not represent a binary phase diagram rather than an isopleth section through the concentration triangle that is shown in Fig.~\ref{fig:triangle}, because every melting/crystallization process is accompanied by mass changes and hence by shifts of the Cu$_2$O/CuO ratio.

Fig.~\ref{fig:triangle}a) describes the system as concentration triangle Al$_2$O$_3$--CuO--Cu. The primary crystallization field of $\alpha$-Al$_2$O$_3$ covers a huge region on top of the triangle, in agreement with Fig.~\ref{fig:PD-section}. Fig.~\ref{fig:triangle}b) is an enlargement of the lowermost part of the triangle, with apex at the position of the star in Fig.~\ref{fig:triangle}a). The small crystallization field of CuAlO$_2$(s) between Cu$_2$O and Al$_2$O$_3$(s) becomes visible which is limited on the right hand side (very reducing conditions) by Cu(s). As shown in Fig.~\ref{fig:predominance} this field is expected to extend more to the left if one takes into account the alloy formation with Pt as a possible fourth component.
At the invariant points 2,3, and 4 the neighboring phases that crystallize first are in equilibrium with the melt. CuAl$_2$O$_4$ spinel can be found only in the vicinity of point 1 but transforms in the subsolidus back to CuAlO$_2$. The spinel can be formed only under oxygen rich conditions where CuO is also stable, and an extended field of primary crystallization of CuAl$_2$O$_4$ was not identified.

\begin{figure}[htb]
\includegraphics[width=0.45\textwidth]{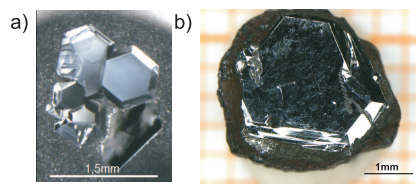}
\caption{CuAlO$_2$ crystals grown from Cu$_2$O solutions with 2\,mol-\% Al$_2$O$_3$ in Pt crucibles in air. a) Spontaneous crystallization in a DTA crucible. b) Spontaneous crystallization in a batch furnace.}
\label{fig:Foto}
\end{figure}

Already in the solidified DTA samples where 2\,mol-\% Al$_2$O$_3$ was added to Cu$_2$O in Pt crucibles, first CuAlO$_2$ crystals were found that are shown in Fig.~\ref{fig:Foto}a). Another test, aimed on checking the long term stability of Pt crucibles against the copper oxide melt was performed in a batch furnace in air. It was found that crucibles with diameters of several centimeters were stable up to seven days. The solidified melt contained CuAlO$_2$ crystals that were formed by spontaneous crystallization, with sizes up to 5\,mm (Fig.~\ref{fig:Foto}b).


\section{Conclusions}

Oxide melts rich in copper oxide can be held stable in platinum crucibles that contain a sufficient oxygen concentration; 15--21\% are the lower limit. If $p_{\mathrm{O}_2}$ is too low, copper oxides are reduced to the metal, because metallic copper is then stabilized by the formation of a solid or liquid solution. CuAlO$_2$ crystallizes first from CuO$_x$ melts containing 1--2\,mol-\% Al$_2$O$_3$, which offers the possibility to grow delafossite crystals from a self-flux without foreign additives. Every melting/crystallization step in this system is connected with a chemical reaction that leads to the absorption or release of oxygen. 


\section*{Acknowledgments}

A. Kwasniewski is thanked for performing X-ray powder analysis and C. Guguschev for performing X-ray fluorescence spectroscopy. We express our gratitude to J. Rehm, I. Schulze-Jonack, E. Thiede and O. Reetz for support in the laboratory, and S. Ganschow for helpful discussions.

The authors acknowledge the funding by the German Research Foundation (DFG) under the project No. SI 463/9-1.



\end{document}